\documentclass[submission]{eptcs}
\providecommand{\event}{TFPIE 2012}
\usepackage{graphicx}
\usepackage{varioref}
\usepackage{url}
\usepackage{syntax}

\makeatletter
\@ifundefined{lhs2tex.lhs2tex.sty.read}%
  {\@namedef{lhs2tex.lhs2tex.sty.read}{}%
   \newcommand\SkipToFmtEnd{}%
   \newcommand\EndFmtInput{}%
   \long\def\SkipToFmtEnd#1\EndFmtInput{}%
  }\SkipToFmtEnd

\newcommand\ReadOnlyOnce[1]{\@ifundefined{#1}{\@namedef{#1}{}}\SkipToFmtEnd}
\usepackage{amstext}
\usepackage{amssymb}
\usepackage{stmaryrd}
\DeclareFontFamily{OT1}{cmtex}{}
\DeclareFontShape{OT1}{cmtex}{m}{n}
  {<5><6><7><8>cmtex8
   <9>cmtex9
   <10><10.95><12><14.4><17.28><20.74><24.88>cmtex10}{}
\DeclareFontShape{OT1}{cmtex}{m}{it}
  {<-> ssub * cmtt/m/it}{}

\DeclareFontShape{OT1}{cmtt}{bx}{n}
  {<5><6><7><8>cmtt8
   <9>cmbtt9
   <10><10.95><12><14.4><17.28><20.74><24.88>cmbtt10}{}
\DeclareFontShape{OT1}{cmtex}{bx}{n}
  {<-> ssub * cmtt/bx/n}{}

\newcommand{\Conid}[1]{\mathit{#1}}
\newcommand{\Varid}[1]{\mathit{#1}}
\newcommand{\anonymous}{\kern0.06em \vbox{\hrule\@width.5em}}
\newcommand{\plus}{\mathbin{+\!\!\!+}}

\usepackage{polytable}

\@ifundefined{mathindent}%
  {\newdimen\mathindent\mathindent\leftmargini}%
  {}%

\def\resethooks{%
  \global\let\SaveRestoreHook\empty
  \global\let\ColumnHook\empty}
\newcommand*{\savecolumns}[1][default]%
  {\g@addto@macro\SaveRestoreHook{\savecolumns[#1]}}
\newcommand*{\restorecolumns}[1][default]%
  {\g@addto@macro\SaveRestoreHook{\restorecolumns[#1]}}
\newcommand*{\aligncolumn}[2]%
  {\g@addto@macro\ColumnHook{\column{#1}{#2}}}

\resethooks

\newcommand{\onelinecommentchars}{\quad-{}- }
\newcommand{\commentbeginchars}{\enskip\{-}
\newcommand{\commentendchars}{-\}\enskip}

\newcommand{\visiblecomments}{%
  \let\onelinecomment=\onelinecommentchars
  \let\commentbegin=\commentbeginchars
  \let\commentend=\commentendchars}

\newcommand{\invisiblecomments}{%
  \let\onelinecomment=\empty
  \let\commentbegin=\empty
  \let\commentend=\empty}

\visiblecomments

\newlength{\blanklineskip}
\newcommand{\hsindent}[1]{\quad}
\let\hspre\empty
\let\hspost\empty

\EndFmtInput
\makeatother
\ReadOnlyOnce{polycode.fmt}%
\makeatletter

\newcommand{\hsnewpar}[1]%
  {{\parskip=0pt\parindent=0pt\par\vskip #1\noindent}}

\newcommand{\hscodestyle}{}

\newcommand{\sethscode}[1]%
  {\expandafter\let\expandafter\hscode\csname #1\endcsname
   \expandafter\let\expandafter\endhscode\csname end#1\endcsname}

  {\par\noindent
   \advance\leftskip\mathindent
   \hscodestyle
   \let\\=\@normalcr
   \let\hspre\(\let\hspost\)%
   \pboxed}%
  {\endpboxed\)%
   \par\noindent
   \ignorespacesafterend}

  {\hsnewpar\abovedisplayskip
   \advance\leftskip\mathindent
   \hscodestyle
   \let\hspre\(\let\hspost\)%
   \pboxed}%
  {\endpboxed%
   \hsnewpar\belowdisplayskip
   \ignorespacesafterend}

  {\hsnewpar\abovedisplayskip
   \advance\leftskip\mathindent
   \hscodestyle
   \let\\=\@normalcr
   \(\pboxed}%
  {\endpboxed\)%
   \hsnewpar\belowdisplayskip
   \ignorespacesafterend}

\newcommand{\plainhs}{\sethscode{plainhscode}}

\plainhs

  {\hsnewpar\abovedisplayskip
   \advance\leftskip\mathindent
   \hscodestyle
   \let\\=\@normalcr
   \(\parray}%
  {\endparray\)%
   \hsnewpar\belowdisplayskip
   \ignorespacesafterend}

  {\parray}{\endparray}

  {\(\parray}{\endparray\)}

\def\codeframewidth{\arrayrulewidth}
\RequirePackage{calc}

  {\parskip=\abovedisplayskip\par\noindent
   \hscodestyle
   \arrayrulewidth=\codeframewidth
   \tabular{@{}|p{\linewidth-2\arraycolsep-2\arrayrulewidth-2pt}|@{}}%
   \hline\framedhslinecorrect\\{-1.5ex}%
   \let\endoflinesave=\\
   \let\\=\@normalcr
   \(\pboxed}%
  {\endpboxed\)%
   \framedhslinecorrect\endoflinesave{.5ex}\hline
   \endtabular
   \parskip=\belowdisplayskip\par\noindent
   \ignorespacesafterend}

\newcommand{\framedhslinecorrect}[2]%
  {#1[#2]}

  {\(\def\column##1##2{}%
   \let\>\undefined\let\<\undefined\let\\\undefined
   \newcommand\>[1][]{}\newcommand\<[1][]{}\newcommand\\[1][]{}%
   \def\fromto##1##2##3{##3}%
   }{\) }%

  {\let\orighscode=\hscode
   \let\origendhscode=\endhscode
   \def\endhscode{\def\hscode{\endgroup\def\@currenvir{hscode}\\}\begingroup}
   \orighscode\def\hscode{\endgroup\def\@currenvir{hscode}}}%
  {\origendhscode
   \global\let\hscode=\orighscode
   \global\let\endhscode=\origendhscode}%

\makeatother
\EndFmtInput
\newcommand{\titlerunning}{Forty hours of declarative programming}
\newcommand{\authorrunning}{J. Stutterheim, W.S. Swierstra \& S.D. Swierstra}

\title{Forty hours of declarative programming\\
\Large{Teaching Prolog at the Junior College Utrecht}}
\author{Jurri\"en Stutterheim \qquad\qquad Wouter Swierstra
\qquad\qquad Doaitse Swierstra
\institute{Department of Information and Computing Sciences\\
Universiteit Utrecht\\
P.O.Box 80.089, 3508 TB Utrecht, The Netherlands}
\email{\{j.stutterheim,w.s.swierstra,doaitse\}@uu.nl}}
\begin{document}
\maketitle

\begin{abstract}
  This paper documents our experience using declarative languages to give
  secondary school students a first taste of Computer Science. The course aims
  to teach students a bit about programming in Prolog, but also exposes them to
  important Computer Science concepts, such as unification or searching
  strategies. Using Haskell's Snap Framework in combination with our own
  \texttt{NanoProlog} library, we have developed a web application to teach
  this course.
\end{abstract}

\section{Introduction}

The Junior College Utrecht (JCU) is a joint initiative of Utrecht University
and 26 local secondary schools. The JCU offers talented students already
interested in science a choice of several short courses at several different
departments of the Faculty of Science of Utrecht University. In their
penultimate year of secondary school---when the students are approximately
seventeen years old---students may do a small research project that encompasses
approximately 40 hours of work, including the preparation of a final report and
a presentation. Students spend two and a half days performing experiments,
followed by two days of preparing their final presentation and report. On the
last afternoon, students present their projects to one another. The courses
consist of five full days, spread over six weeks. The students may indicate
their preference for several different topics, including biology, veterinary
science, chemistry, or mathematics. Finally, the students are distributed over
the available courses, taking their preferences and the available space on each
course into account. In the past two years, we have developed a Computer
Science course for these students~\cite{swierstra2011}.

Choosing suitable material for such a course is not easy.  Part of the purpose
of the JCU is to ignite the students' interest in (Computer) Science. While we
want the students to learn a bit of computer programming, the course should
strive to teach more. We would like our students to learn something about
algorithms, semantics, programming language design, and other abstract concepts
from Computer Science. In our experience, very few students who start the
course have a clear idea of what Computer Science is. One of our main goals was
to ensure that students that complete our course learn what Computer Science is
and how it differs from Mathematics and other Sciences. Finally, we hope to
encourage students to enrol at Utrecht University---above all else, the course
should be fun.

Early on, we decided that we wanted to teach the JCU students to program.
Seeing your own code execute on your computer for the first time is a magical
experience. Suddenly students become aware that they can instruct the computer
to do what they tell it. Instead of simply \emph{consuming} the applications
others have written, students learn that a computer is a \emph{programmable}
device that they can control themselves.

Once we decided that we wanted to teach a programming language, the next
question is of course \emph{which} language to teach. Should we teach a
mainstream programming language, such as Java or C\#? Or should we teach a
`toy' language such as Alice~\cite{conway1997} or Scratch~\cite{maloney2004}?
Or should we teach languages popular in the functional programming language
research community such as Agda~\cite{norell}, Epigram~\cite{view},
OCaml~\cite{ocaml}, Racket~\cite{racket}, or Haskell~\cite{haskell98}?  We
opted for none of the above.

The following requirements helped us to determine our choice of language:

\begin{itemize}
\item The language needs to be \emph{simple}. Students should be up and running
  as quickly as possible. As we only have approximately two and a half days to
  teach the students, we want a language with minimal opportunity for making
  syntax errors or type errors, and getting incomprehensible compiler messages
  as a result.
\item The language needs to be non-trivial. Many introductory programming
  languages target a younger, pre-teen audience by offering drag \& drop IDEs
  and focussing on drawing and/or animating images.  The language should be
  challenging enough to intrigue clever, seventeen year-old secondary school
  students. A language like Scratch, for example, is great for simple
  animations and games, but it does not illustrate what real Computer Science
  is about and how real-world problems can be solved by programming.
\item The language needs to create a level playing field. Some students already
  have programming experience when starting the JCU course. This can be
  extremely discouraging for students without prior programming experience.
  Inexperienced students may get the (incorrect) impression that they are not
  suitable for a degree in Computer Science.
\item The students have no experience with Computer Science. The Mathematics
  course taught at their secondary schools, where they learn some elementary
  calculus and geometry, is the closest related subject. To engage the
  students, the language we choose should relate to what they already know.
\end{itemize}

In the end, we settled upon (a fragment of) Prolog. This may seem like a
peculiar choice, so we would like to provide some further motivation for this
decision:

\begin{itemize}
\item The students have no trouble learning the syntax. Typically, they are
  writing their first Prolog relations within hours. Had we tried to teach
  Java, for example, we would need a significant amount of time to explain what
  \texttt{public static void} means and how to deal with curly braces and
  semi-colons. Choosing Prolog avoids these problems altogether.

\item They can solve non-trivial problems by writing out a precise,
  mathematical specification. In our course, for instance, the students develop
  a small Sudoku solver. After the students have developed a specification of
  what a valid solution to a Sudoku problem is, Prologs search automatically
  provides them with a Sudoku \emph{solver}. In the remainder of the course,
  the students can cover a wide variety of problem domains, including
  scheduling problems, routing problems, basic genetics, and logical
  brainteasers. In each of these domains, they learn how to formalize some
  abstract concept, resulting in an executable program.

\item Using Prolog gives us the opportunity to teach the students about
  backtracking, unification, proof trees, and many other important concepts. We
  hope that by doing so, they learn there is more to Computer Science than
  `just programming.'

\item We can teach them how an interpreter works. We have implemented a minimal
  interpreter for Prolog in Haskell, and use this to teach our students a bit
  about functional programming. Not all students study this material, but the
  students with previous programming experience find this by far the most
  interesting part of the course.

\item Prolog has no type system. While reasoning about types is an important
  aspect in the functional programming courses at Utrecht University, teaching
  the JCU students about types and have them write interesting programs within
  40 hours seems unrealistic. In Prolog, the students can easily write an run
  their programs without having to decipher type-error messages first.

\end{itemize}

We have developed a web application to support this course. The application
consists of two modes of operation: a web-based interface to our Prolog
interpreter and an interactive Prolog proof assistant. This application,
including its browser-based front-end, was built entirely in Haskell using the
Snap web framework and the Utrecht Haskell Compiler (UHC). In the following
sections we will discuss the structure of our course
(Section~\ref{sec:coursesetup}), the implementation of our Prolog interpreter
(Section~\ref{sec:nanoprolog}), the interactive proof assistant
(Section~\ref{sec:web-app}), and the students' experience working on this
course, using these tools (Section~\ref{sec:discussion}).

\section{Course setup}
\label{sec:coursesetup}

\subsection{Planning}
\label{sec:planning}

As the title of this paper suggests, we only have a very limited amount of time
to teach this course. Unfortunately, this time constraint is put in place by
the JCU itself, and is not something we as teachers of an individual pre-thesis
track can easily change. In practice, we only have five days of about 6 to 7.5
hours each to introduce the students to Prolog, and assist them in doing their
research project, preparing their presentation, and writing their report. The
entire course is roughly structured as follows:

\paragraph{Day 1} We meet the students for the first time. After a short
introduction, the students form teams of preferably two, but if need be
three people. With the teams formed, they start reading our lecture
notes~\cite{swierstra2011} and start working on the exercises we have prepared.

The exercises start off slowly with an introduction to substitutions. Afterwards, the lecture notes continue
with an introduction to Prolog and depth-first search. The lecture notes
conclude with an explanation of how our Haskell implementation of a Prolog
interpreter works. Before students can study this chapter, they are referred to
the first two chapters of Learn You a Haskell~\cite{lipovaca2011}.  In
practice, very few of students get around to the Haskell part of the course.

\paragraph{Day 2} The students get some time to finish working on the lecture
notes and ask questions about it. Afterwards, each team selects a research
project and starts working on it. Students can come up with their own project,
but since they generally cannot correctly determine the scope of a project
themselves, we present them with a list of projects from which to choose.


\paragraph{Day 3} Work continues on the research project and the students start
working on a report in which they detail what they have learnt during the
course. This report is mostly aimed at the course's teachers and is expected to
discuss the technical implementation details of their research project. In
addition, the students are asked to reflect on how they executed their research
and how they might improve in a future project.

\paragraph{Day 4} In addition to continuing work on their report, the students
start working on their presentation. The presentation needs to be aimed at a
general audience of parents, teachers from other disciplines, and fellow
students that have chosen other topics at the JCU. In practice, the students
utilize this day to continue work on their research project as well.

\paragraph{Day 5} At the end of the day, the students will need to present
their projects to a general audience of both parents and teachers, hence this
day is mostly concerned with practising and improving the presentations.

One week after the final presentations, the students have to hand in their
report.

\subsection{Course material}
\label{sec:course-material}

The lecture notes provided to the students provides them with an
introduction to Prolog and its implementation. The course notes start
with a short introduction on substitutions, equations, and variables
in terms of the high-school mathematics with which the students are
already familiar. On the surface, this has little to do with
programming, but it aims to fix some terminology before introducing
Prolog.

Next, they learn a bit about Prolog through a series of examples,
starting with a set of relations to describe a family tree. The first
queries they write themselves are along the lines of `Who are the
children of Alice?' or `Who are Bob's grandparents?'. By practicing
with these simple examples, they gain some basic familiarity with
Prolog, but also learn the difference between existential and
universal quantification. To understand simple rules such as:
\begin{hscode}\SaveRestoreHook
\column{B}{@{}>{\hspre}l<{\hspost}@{}}%
\column{3}{@{}>{\hspre}l<{\hspost}@{}}%
\column{E}{@{}>{\hspre}l<{\hspost}@{}}%
\>[3]{}\Varid{grandparent}\;(\Conid{X},\Conid{Y}):\!\text{-}\,\Varid{parent}\;(\Conid{X},\Conid{Z}),\Varid{parent}\;(\Conid{Z},\Conid{Y}).{}\<[E]%
\ColumnHook
\end{hscode}\resethooks
You already need to understand that \ensuremath{\Conid{X}} and \ensuremath{\Conid{Y}} are universally
quantified and that \ensuremath{\Conid{Z}} is existentially quantified. 

We introduce \emph{recursion} very early on, starting with a
definition of an \ensuremath{\Varid{ancestors}} relation. We then move on to simple Peano
arithmetic, using recursion to define addition and multiplication.

Once they have seen Peano arithmetic, the next step is to study simple
list processing. We show how to define simple functions such as
\ensuremath{\Varid{length}} and \ensuremath{\Varid{append}} in Prolog. This all builds up to a (mini) Sudoku
solver. This Sudoku solver itself merely formally describes what a solution is,
leaving the search to Prolog's resolution mechanism. We start by
defining predicates that state that a given list contains no
duplicates. Using this, we define that all the rows and columns of a
given board have no duplicates. It is then up to the students to
complete the solver, by defining what it means for the `boxes' of a
Sudoku puzzle to have no duplicates.

Having completed the Sudoku solver, the course notes shift towards
describing how Prolog resolution works. At first, we present a
pseudo-code algorithm, exposing the students to concepts such as
\emph{unification}, \emph{search trees}, and \emph{depth-first
  search}. The final chapter of the notes explains how this
pseudo-code can be implemented in Haskell. 

\subsection{Projects}
\label{sec:projects}

Once the students have worked their way through the lecture notes,
they start their own short research project. The research projects
take the form of a more open ended challenge. Many students choose to
use the Prolog interpreter already present on the lab machines, rather
than use the web interface. This allows them to use a regular text
editor to write their programs and use some of the features of Prolog
that our simple interpreter does not support, such as special syntax
for natural numbers. 

The projects they work on are quite varied. Some are inspired by
Sudoku-like puzzles, such as Kakuro. Others build on the family tree
example, asking them to define new family relations or to extend the
relations with some basic genetics (e.g., if both your parents exhibit
the same recessive trait, you share that trait). Furthermore, we offer
them a choice of several other Prolog programming task, such as
defining a toy route planner system or simple scheduling
software. Finally, we suggest several projects that modify the
interpreter, such as implementing a breadth-first search or adding a
cut operator. These last exercises tend to be quite daunting.

To give you a better idea of the kind of project the students write,
we will briefly discuss the routing software that one team
wrote last year. They started with having to define a binary relation
between a fixed number of cities that represents that two cities are
immediately connected. They then extended this to a recursive relation
that holds if two cities are connected by zero or more steps. This is a
bit subtle: you need to keep track of a list of cities that you have
already visited to ensure that you do not loop. Finally, we asked the
students to extend this simple algorithm to also compute the shortest
path using iterative deepening.

All in all, the students tend to write a small amount of code; a project is
typically between 30 to 100 lines of Prolog. In the bigger programs,
a large part of the code often consists of enumerations of various base cases.
The students find it challenging to have to start from scratch, but at the same
time they enjoy the creative freedom associated with programming.

\section{NanoProlog Interpreter}
\label{sec:nanoprolog}

To allow students to play with Prolog, and--for the really fast learners--to
learn how a Prolog interpreter works, we have written a small Prolog
interpreter of our own, called \textit{NanoProlog}. The core of our interpreter
consists of 150 lines of Haskell, excluding the parser and main I/O loop. This
section aims to give you an idea of how our NanoProlog interpreter works. The
complete source code is available from Hackage as the \texttt{NanoProlog}
package. We hope that it is simple enough that motivated students, with our
help, will be able to understand how this interpreter implements Prolog's
backtracking search.

The central data types we use to represent Prolog programs are defined as
follows:

\begin{hscode}\SaveRestoreHook
\column{B}{@{}>{\hspre}l<{\hspost}@{}}%
\column{3}{@{}>{\hspre}l<{\hspost}@{}}%
\column{5}{@{}>{\hspre}c<{\hspost}@{}}%
\column{5E}{@{}l@{}}%
\column{8}{@{}>{\hspre}l<{\hspost}@{}}%
\column{E}{@{}>{\hspre}l<{\hspost}@{}}%
\>[3]{}\mathbf{data}\;\Conid{Term}{}\<[E]%
\\
\>[3]{}\hsindent{2}{}\<[5]%
\>[5]{}\mathrel{=}{}\<[5E]%
\>[8]{}\Conid{Var}\;\Conid{String}{}\<[E]%
\\
\>[3]{}\hsindent{2}{}\<[5]%
\>[5]{}\mid {}\<[5E]%
\>[8]{}\Conid{Fun}\;\Conid{String}\;[\mskip1.5mu \Conid{Term}\mskip1.5mu]{}\<[E]%
\\[\blanklineskip]%
\>[3]{}\mathbf{data}\;\Conid{Rule}\mathrel{=}\Conid{Term}:\!\leftarrow{}\!\!\!\!\!:[\mskip1.5mu \Conid{Term}\mskip1.5mu]{}\<[E]%
\\[\blanklineskip]%
\>[3]{}\mathbf{type}\;\Conid{Program}\mathrel{=}[\mskip1.5mu \Conid{Rule}\mskip1.5mu]{}\<[E]%
\ColumnHook
\end{hscode}\resethooks
A Prolog \ensuremath{\Conid{Term}} is either a variable or a function constant applied to a list
of terms. To simplify parsing, we require all variables to start with a capital
letter; all constants should start with a lower-case letter. A \ensuremath{\Conid{Rule}} describes
a single Prolog inference rule. The rule \ensuremath{\Varid{t}:\!\leftarrow{}\!\!\!\!\!:\Varid{ts}} states that to prove the
goal \ensuremath{\Varid{t}}, it is sufficient to prove each sub-goal in the list \ensuremath{\Varid{ts}}. For
example, a Prolog rule such as
\begin{hscode}\SaveRestoreHook
\column{B}{@{}>{\hspre}l<{\hspost}@{}}%
\column{3}{@{}>{\hspre}l<{\hspost}@{}}%
\column{E}{@{}>{\hspre}l<{\hspost}@{}}%
\>[3]{}\Varid{ancestor}\;(\Conid{X},\Conid{Y}):\!\text{-}\,\Varid{parent}\;(\Conid{Z},\Conid{Y}),\Varid{ancestor}\;(\Conid{X},\Conid{Z}){}\<[E]%
\ColumnHook
\end{hscode}\resethooks
will be parsed as:
\begin{hscode}\SaveRestoreHook
\column{B}{@{}>{\hspre}l<{\hspost}@{}}%
\column{3}{@{}>{\hspre}l<{\hspost}@{}}%
\column{4}{@{}>{\hspre}l<{\hspost}@{}}%
\column{E}{@{}>{\hspre}l<{\hspost}@{}}%
\>[3]{}(\Conid{Fun}\;\text{\tt \char34 ancestor\char34}\;[\mskip1.5mu \Conid{Var}\;\text{\tt \char34 X\char34},\Conid{Var}\;\text{\tt \char34 Y\char34}\mskip1.5mu]):\!\leftarrow{}\!\!\!\!\!:{}\<[E]%
\\
\>[3]{}\hsindent{1}{}\<[4]%
\>[4]{}[\mskip1.5mu \Conid{Fun}\;\text{\tt \char34 parent\char34}\;[\mskip1.5mu \Conid{Var}\;\text{\tt \char34 Z\char34},\Conid{Var}\;\text{\tt \char34 Y\char34}\mskip1.5mu]{}\<[E]%
\\
\>[3]{}\hsindent{1}{}\<[4]%
\>[4]{},\Conid{Fun}\;\text{\tt \char34 ancestor\char34}\;[\mskip1.5mu \Conid{Var}\;\text{\tt \char34 Z\char34},\Conid{Var}\;\text{\tt \char34 Y\char34}\mskip1.5mu]\mskip1.5mu]{}\<[E]%
\ColumnHook
\end{hscode}\resethooks
Finally, a program consists of a list of such rules.

The Prolog interpreter takes a program as input, together with a goal query. It
constructs an inhabitant of the following \ensuremath{\Conid{Result}} type:
\begin{hscode}\SaveRestoreHook
\column{B}{@{}>{\hspre}l<{\hspost}@{}}%
\column{3}{@{}>{\hspre}l<{\hspost}@{}}%
\column{E}{@{}>{\hspre}l<{\hspost}@{}}%
\>[3]{}\mathbf{newtype}\;\Conid{Env}\mathrel{=}\Conid{Env}\;\{\mskip1.5mu \Varid{fromEnv}\mathbin{::}\Conid{Map}\;\Conid{String}\;\Conid{Term}\mskip1.5mu\}{}\<[E]%
\ColumnHook
\end{hscode}\resethooks
\begin{hscode}\SaveRestoreHook
\column{B}{@{}>{\hspre}l<{\hspost}@{}}%
\column{3}{@{}>{\hspre}l<{\hspost}@{}}%
\column{5}{@{}>{\hspre}c<{\hspost}@{}}%
\column{5E}{@{}l@{}}%
\column{8}{@{}>{\hspre}l<{\hspost}@{}}%
\column{E}{@{}>{\hspre}l<{\hspost}@{}}%
\>[3]{}\mathbf{data}\;\Conid{Result}{}\<[E]%
\\
\>[3]{}\hsindent{2}{}\<[5]%
\>[5]{}\mathrel{=}{}\<[5E]%
\>[8]{}\Conid{Done}\;\Conid{Env}{}\<[E]%
\\
\>[3]{}\hsindent{2}{}\<[5]%
\>[5]{}\mid {}\<[5E]%
\>[8]{}\Conid{Apply}\;[\mskip1.5mu \Conid{Result}\mskip1.5mu]{}\<[E]%
\ColumnHook
\end{hscode}\resethooks
The answers we are looking for are substitutions, mapping variables to Prolog
terms. We represent such substitutions as environments, using Haskell's
\ensuremath{\Conid{\Conid{Data}.Map}} library. If the answer to our query is simple enough, there might
not be any variables and the environment may be empty.  The \ensuremath{\Conid{Result}} data type
represents the search tree that is built up during resolution. It has two
constructors: the \ensuremath{\Conid{Done}} constructor returns the required solution; the \ensuremath{\Conid{Apply}}
branches over all possible rules that can be applied to solve the current goal.

The resolution process now proceeds in two steps: we begin by constructing a
\ensuremath{\Conid{Result}} data type. By traversing this tree in any order, we can search for
answers.

The \ensuremath{\Varid{solve}} function below forms the heart of our interpreter. The function
unifies the current goal with the conclusion of every possible rule. When
unification succeeds, we proceed by resolving the right-hand side of the rule,
together with any remaining goals. When an \ensuremath{\Conid{Apply}} node does not have any
children, this represents a `dead-end' with no possible solutions.
\begin{hscode}\SaveRestoreHook
\column{B}{@{}>{\hspre}l<{\hspost}@{}}%
\column{3}{@{}>{\hspre}l<{\hspost}@{}}%
\column{5}{@{}>{\hspre}l<{\hspost}@{}}%
\column{7}{@{}>{\hspre}l<{\hspost}@{}}%
\column{16}{@{}>{\hspre}l<{\hspost}@{}}%
\column{24}{@{}>{\hspre}l<{\hspost}@{}}%
\column{54}{@{}>{\hspre}c<{\hspost}@{}}%
\column{54E}{@{}l@{}}%
\column{57}{@{}>{\hspre}l<{\hspost}@{}}%
\column{68}{@{}>{\hspre}l<{\hspost}@{}}%
\column{E}{@{}>{\hspre}l<{\hspost}@{}}%
\>[3]{}\mathbf{type}\;\Conid{Goal}\mathrel{=}\Conid{Term}{}\<[E]%
\\[\blanklineskip]%
\>[3]{}\Varid{solve}\mathbin{::}\Conid{Program}\to \Conid{Goal}\to \Conid{Result}{}\<[E]%
\\
\>[3]{}\Varid{solve}\;\Varid{rules}\;\Varid{goal}\mathrel{=}\Varid{steps}\;(\Conid{Env}\;\Varid{empty})\;[\mskip1.5mu \Varid{goal}\mskip1.5mu]{}\<[E]%
\\
\>[3]{}\hsindent{2}{}\<[5]%
\>[5]{}\mathbf{where}{}\<[E]%
\\
\>[5]{}\hsindent{2}{}\<[7]%
\>[7]{}\Varid{steps}\mathbin{::}\Conid{Env}\to [\mskip1.5mu \Conid{Goal}\mskip1.5mu]\to \Conid{Result}{}\<[E]%
\\
\>[5]{}\hsindent{2}{}\<[7]%
\>[7]{}\Varid{steps}\;\Varid{e}\;{}\<[16]%
\>[16]{}[\mskip1.5mu \mskip1.5mu]{}\<[24]%
\>[24]{}\mathrel{=}\Conid{Done}\;\Varid{e}{}\<[E]%
\\
\>[5]{}\hsindent{2}{}\<[7]%
\>[7]{}\Varid{steps}\;\Varid{e}\;{}\<[16]%
\>[16]{}(\Varid{g}\mathbin{:}\Varid{gs}){}\<[24]%
\>[24]{}\mathrel{=}\Conid{Apply}\;[\mskip1.5mu \Varid{steps}\;\Varid{e'}\;(\Varid{cs}\plus \Varid{gs}){}\<[54]%
\>[54]{}\mid {}\<[54E]%
\>[57]{}\Varid{c}:\!\leftarrow{}\!\!\!\!\!:\Varid{cs}{}\<[68]%
\>[68]{}\leftarrow \Varid{rules}{}\<[E]%
\\
\>[54]{},{}\<[54E]%
\>[57]{}\Conid{Just}\;\Varid{e'}{}\<[68]%
\>[68]{}\leftarrow [\mskip1.5mu \Varid{unify}\;(\Varid{g},\Varid{c})\;(\Conid{Just}\;\Varid{e})\mskip1.5mu]\mskip1.5mu]{}\<[E]%
\ColumnHook
\end{hscode}\resethooks
The actual implementation also tracks which rules are applied at every step, so
that we can not only return the successful substitutions, but also the trace of
all the rules that were applied to construct this solution.

The \ensuremath{\Varid{unify}} code is shown below. When the \ensuremath{\Varid{unify}} function is applied to two
terms and a non-empty environment, the two terms are unified after the
variables in the terms have been substituted by values from the environment. If
one of the two terms is a variable, a substitution from the variable to the
other term is inserted in the environment. If both terms are a \ensuremath{\Conid{Fun}}, the
right-hand sides of the rules are unified if and only if both rules have the
same name and the same number of terms on the right-hand side.
\begin{hscode}\SaveRestoreHook
\column{B}{@{}>{\hspre}l<{\hspost}@{}}%
\column{3}{@{}>{\hspre}l<{\hspost}@{}}%
\column{5}{@{}>{\hspre}l<{\hspost}@{}}%
\column{7}{@{}>{\hspre}l<{\hspost}@{}}%
\column{9}{@{}>{\hspre}c<{\hspost}@{}}%
\column{9E}{@{}l@{}}%
\column{12}{@{}>{\hspre}l<{\hspost}@{}}%
\column{17}{@{}>{\hspre}l<{\hspost}@{}}%
\column{24}{@{}>{\hspre}l<{\hspost}@{}}%
\column{39}{@{}>{\hspre}l<{\hspost}@{}}%
\column{46}{@{}>{\hspre}l<{\hspost}@{}}%
\column{54}{@{}>{\hspre}l<{\hspost}@{}}%
\column{70}{@{}>{\hspre}l<{\hspost}@{}}%
\column{73}{@{}>{\hspre}l<{\hspost}@{}}%
\column{E}{@{}>{\hspre}l<{\hspost}@{}}%
\>[3]{}\Varid{unify}\mathbin{::}(\Conid{Term},\Conid{Term})\to \Conid{Maybe}\;\Conid{Env}\to \Conid{Maybe}\;\Conid{Env}{}\<[E]%
\\
\>[3]{}\Varid{unify}\;\anonymous \;{}\<[17]%
\>[17]{}\Conid{Nothing}{}\<[39]%
\>[39]{}\mathrel{=}\Conid{Nothing}{}\<[E]%
\\
\>[3]{}\Varid{unify}\;(\Varid{t},\Varid{u})\;{}\<[17]%
\>[17]{}\Varid{env}\mathord{@}(\Conid{Just}\;\Varid{e}\mathord{@}(\Conid{Env}\;\Varid{m})){}\<[39]%
\>[39]{}\mathrel{=}\Varid{uni}\;(\Varid{subst}\;\Varid{e}\;\Varid{t})\;(\Varid{subst}\;\Varid{e}\;\Varid{u}){}\<[E]%
\\
\>[3]{}\hsindent{2}{}\<[5]%
\>[5]{}\mathbf{where}{}\<[E]%
\\
\>[5]{}\hsindent{2}{}\<[7]%
\>[7]{}\Varid{uni}\;{}\<[12]%
\>[12]{}(\Conid{Var}\;\Varid{x})\;{}\<[24]%
\>[24]{}\Varid{y}{}\<[46]%
\>[46]{}\mathrel{=}\Conid{Just}\;{}\<[54]%
\>[54]{}(\Conid{Env}\;(\Varid{insert}\;\Varid{x}\;{}\<[70]%
\>[70]{}\Varid{y}\;{}\<[73]%
\>[73]{}\Varid{m})){}\<[E]%
\\
\>[5]{}\hsindent{2}{}\<[7]%
\>[7]{}\Varid{uni}\;{}\<[12]%
\>[12]{}\Varid{x}\;{}\<[24]%
\>[24]{}(\Conid{Var}\;\Varid{y}){}\<[46]%
\>[46]{}\mathrel{=}\Conid{Just}\;{}\<[54]%
\>[54]{}(\Conid{Env}\;(\Varid{insert}\;\Varid{y}\;{}\<[70]%
\>[70]{}\Varid{x}\;{}\<[73]%
\>[73]{}\Varid{m})){}\<[E]%
\\
\>[5]{}\hsindent{2}{}\<[7]%
\>[7]{}\Varid{uni}\;{}\<[12]%
\>[12]{}(\Conid{Fun}\;\Varid{x}\;\Varid{xs})\;{}\<[24]%
\>[24]{}(\Conid{Fun}\;\Varid{y}\;\Varid{ys}){}\<[E]%
\\
\>[7]{}\hsindent{2}{}\<[9]%
\>[9]{}\mid {}\<[9E]%
\>[12]{}\Varid{x}\equiv \Varid{y}\mathrel{\wedge}\Varid{length}\;\Varid{xs}\equiv \Varid{length}\;\Varid{ys}{}\<[46]%
\>[46]{}\mathrel{=}\Varid{foldr}\;\Varid{unify}\;\Varid{env}\;(\Varid{zip}\;\Varid{xs}\;\Varid{ys}){}\<[E]%
\\
\>[5]{}\hsindent{2}{}\<[7]%
\>[7]{}\Varid{uni}\;{}\<[12]%
\>[12]{}\anonymous \;{}\<[24]%
\>[24]{}\anonymous {}\<[46]%
\>[46]{}\mathrel{=}\Conid{Nothing}{}\<[E]%
\ColumnHook
\end{hscode}\resethooks
The \ensuremath{\Conid{Subst}} typeclass is used to overload the \ensuremath{\Varid{subst}} function to work on
terms, lists of terms, and rules.
\begin{hscode}\SaveRestoreHook
\column{B}{@{}>{\hspre}l<{\hspost}@{}}%
\column{3}{@{}>{\hspre}l<{\hspost}@{}}%
\column{5}{@{}>{\hspre}l<{\hspost}@{}}%
\column{27}{@{}>{\hspre}l<{\hspost}@{}}%
\column{E}{@{}>{\hspre}l<{\hspost}@{}}%
\>[3]{}\mathbf{class}\;\Conid{Subst}\;\Varid{t}\;\mathbf{where}{}\<[E]%
\\
\>[3]{}\hsindent{2}{}\<[5]%
\>[5]{}\Varid{subst}\mathbin{::}\Conid{Env}\to \Varid{t}\to \Varid{t}{}\<[E]%
\\[\blanklineskip]%
\>[3]{}\mathbf{instance}\;\Conid{Subst}\;\Varid{a}\Rightarrow \Conid{Subst}\;[\mskip1.5mu \Varid{a}\mskip1.5mu]\;\mathbf{where}{}\<[E]%
\\
\>[3]{}\hsindent{2}{}\<[5]%
\>[5]{}\Varid{subst}\;\Varid{env}\mathrel{=}\Varid{map}\;(\Varid{subst}\;\Varid{env}){}\<[E]%
\\[\blanklineskip]%
\>[3]{}\mathbf{instance}\;\Conid{Subst}\;\Conid{Term}\;\mathbf{where}{}\<[E]%
\\
\>[3]{}\hsindent{2}{}\<[5]%
\>[5]{}\Varid{subst}\;\Varid{env}\;(\Conid{Var}\;\Varid{x}){}\<[27]%
\>[27]{}\mathrel{=}\Varid{maybe}\;(\Conid{Var}\;\Varid{x})\;(\Varid{subst}\;\Varid{env})\;(\Varid{lookup}\;\Varid{x}\;(\Varid{fromEnv}\;\Varid{env})){}\<[E]%
\\
\>[3]{}\hsindent{2}{}\<[5]%
\>[5]{}\Varid{subst}\;\Varid{env}\;(\Conid{Fun}\;\Varid{x}\;\Varid{cs}){}\<[27]%
\>[27]{}\mathrel{=}\Conid{Fun}\;\Varid{x}\;(\Varid{subst}\;\Varid{env}\;\Varid{cs}){}\<[E]%
\\[\blanklineskip]%
\>[3]{}\mathbf{instance}\;\Conid{Subst}\;\Conid{Rule}\;\mathbf{where}{}\<[E]%
\\
\>[3]{}\hsindent{2}{}\<[5]%
\>[5]{}\Varid{subst}\;\Varid{env}\;(\Varid{c}:\!\leftarrow{}\!\!\!\!\!:\Varid{cs})\mathrel{=}\Varid{subst}\;\Varid{env}\;\Varid{c}:\!\leftarrow{}\!\!\!\!\!:\Varid{subst}\;\Varid{env}\;\Varid{cs}{}\<[E]%
\ColumnHook
\end{hscode}\resethooks
Our NanoProlog interpreter does have several restrictions. There is no \ensuremath{\Varid{cut}}
operator; there is no way to define negation; there are no built-in integers,
strings, lists, IO, or other functionality. To keep the implementation minimal,
there is no occurs check. Yet the resulting, simple language is just enough to
solve simple, but interesting problems.











\section{Web Application}
\label{sec:web-app}
To allow the students to experiment with Prolog, without having to install any
software themselves, we have developed a web application. The application has
two modes of operation: as an interactive `theorem prover' and as a Prolog
interpreter.

In the first mode, we aim to teach the students how unification and
backtracking work. Students can enter a Prolog query and then try to `prove' it
themselves by dragging and dropping rules from a list onto the query, thereby
constructing a proof tree.  Dropping a rule on a term unifies the conclusion of
the rule with that term. The body of the rule may then introduce new sub-goals,
expanding the proof tree. This is repeated until all the proof tree's leaves
are basic Prolog facts, and the proof is complete.

\begin{figure}[hbtp]
  \centering
  \includegraphics[height=\textheight]{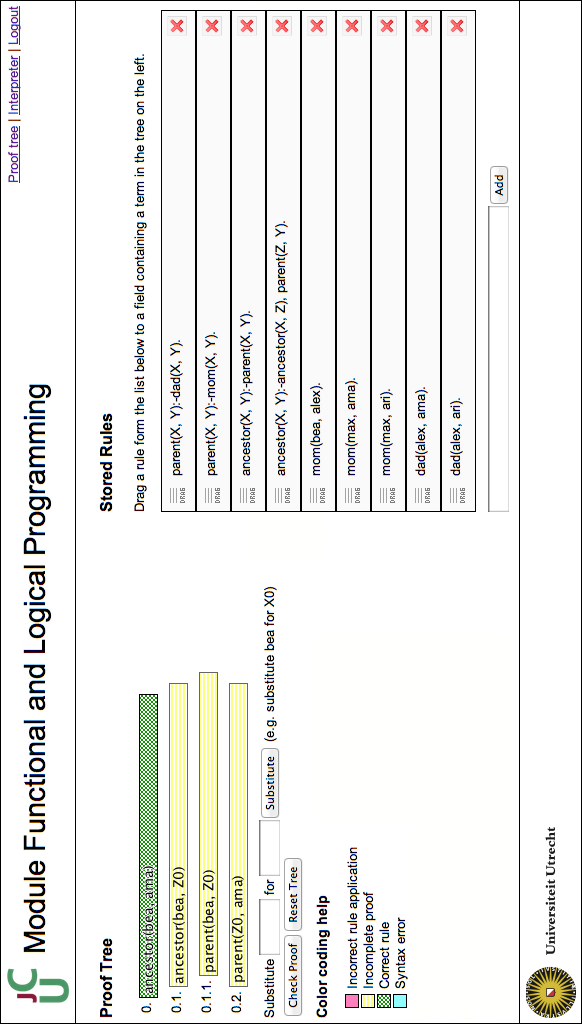}
  \caption{An incomplete proof}
  \label{fig:incomplete-proof}
\end{figure}

A student can ask for feedback while working on the proof. When a branch of the
tree still has open sub-goals, its leaves are rendered with a yellow
background, indicating that there is still some work to be done. Nodes that
have been successfully completed turn green. When all nodes have turned green,
the proof is complete and the student is congratulated with a message box.
Figure \vref{fig:incomplete-proof} shows an incomplete proof. The root node has
successfully been proved, but its children still contain open goals. By
dragging and dropping the rules, listed on the right, on the open sub-goals the
students can complete the proof.

Variables can be substituted in two ways. When a rule's conclusion is unified
with an open sub goal, a substitution is produced and applied to the entire
tree. Alternatively, the students can manually apply a substitution by using
the controls just below the proof tree.

\begin{figure}[hbtp]
  \centering
  \includegraphics[height=\textheight]{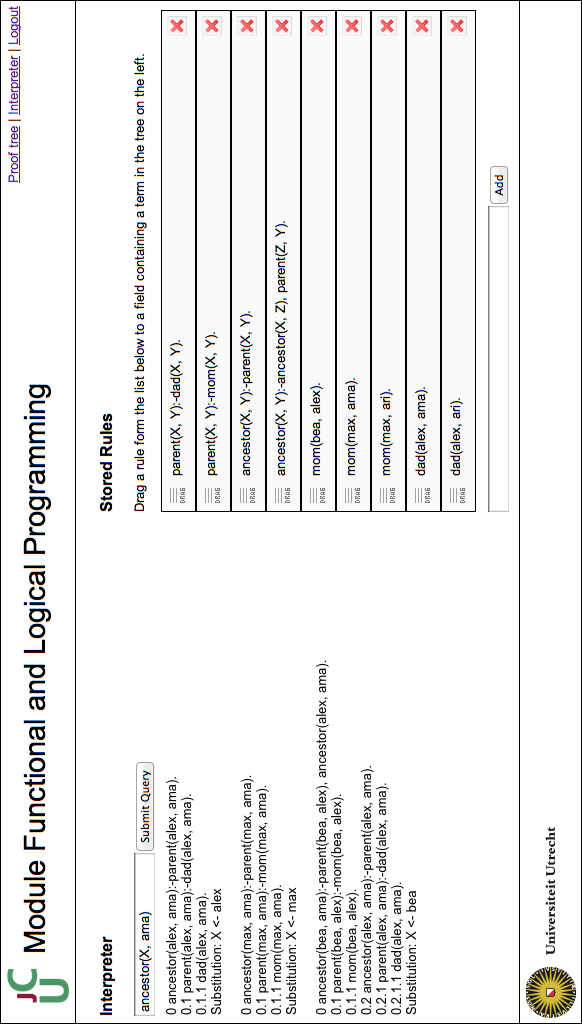}
  \caption{A proof found by the web-based interpreter}
  \label{fig:interpreter-proof}
\end{figure}

The second way the students can use the application is by using it as an
interpreter. Figure~\vref{fig:interpreter-proof} illustrates this mode of
development. Students are presented with a single text field in which they can
enter a Prolog query. If the query can be proven, using the rules on the right,
the interpreter will print the corresponding proof trees and substitutions on
the screen. Students can define new rules, using the text field at the
bottom-right.

Before being able to use the application, students need to register. This
enables us to store a set of rules for each individual student, giving them
their own playground to experiment with Prolog. Having a web application that
the students can access from home is a huge advantage. The students do not need
to install their software and can experiment freely with our tools. For many of
the larger projects the students do use SWI Prolog, that is already installed
on the university computers.

\subsection*{Implementation}
We have implemented both the client and the server part of the application
entirely in Haskell. On the server side, we use the Snap Framework~\cite{snap}
to expose a RESTful~\cite{fielding2000} API to the NanoProlog interpreter and a
database in which we store the Prolog rules the user has defined.

On the client side, we make use of the Utrecht Haskell Compiler's
(UHC)~\cite{www09uhc,dijkstra09uhc-arch} JavaScript
back-end~\cite{dijkstra2012}. This enables us to write Haskell code and
cross-compile it to JavaScript, which is run in the students' browser. Earlier
revisions of the software were implemented in plain JavaScript and delegated
parsing terms, unification and proof verification to the NanoProlog library on
the server. By compiling the NanoProlog library to JavaScript, these operations
can be executed client-side, eliminating additional AJAX requests. The `theorem
prover' part of the application---the first mode of operation---is implemented
completely in Haskell and is running entirely on the client. Unfortunately, the
JavaScript code generated by the UHC is not fast enough to run the graphical
user interface to the interpreter---the second mode of operation---on the
client completely. Instead, the user interface delegates the queries entered by
the user to the server, where the much faster native code interprets them and
sends the results back to the client.

\section{Discussion}
\label{sec:discussion}

\subsection*{Results}
\label{sec:results}

As we mentioned in Section~\ref{sec:coursesetup}, we only have about two and a
half days to teach the students. The remaining time is reserved for the
preparation of a presentation, the writing of a final report, and a
mini-symposium where all students across the different sciences present their
results.

The first year we ran this course, it was quite `free-form', in part because we
had not assembled all the required material. Instead of spoon-feeding the
students, we hoped that they would be capable of some individual exploration:
the JCU specifically aims to teach students how scientific research works. In
practice, however, these students, regardless of how motivated they are, are
too young to work independently.

The second time we ran the course, we provided the students much more
structure. We had prepared a set of course notes with numerous small exercises.
This gave the students a clear initial goal: read the course notes, complete
the exercises, and learn to use the web application. After completing the
notes, the students embarked on their research projects.
This gave the students the freedom to work on a small research project, but
still provides enough guidance.

\paragraph{Student Reflection}

Throughout the course, the students were under constant supervision, which
allowed them to frequently ask questions. While working through the course
notes, the concept that the students seem to struggle with most was the scope
of variables. The students found it hard to keep in mind that a variable \ensuremath{\Conid{X}} in
one rule is not the same variable \ensuremath{\Conid{X}} in another rule, especially in the case
of recursion. Luckily, by the time  they moved on to their projects, this
confusion had nearly completely disappeared.

At the end of their research project, all of the students are required
to reflect on their research by writing a short report. We summarize
the experiences of the spring 2011 class here, which consisted of
eleven students. Some of them had some prior programming experience,
but most did not.

Especially the students without programming experience indicated that
they did not really know what Computer Science was or how difficult
the course was going to be. Some even indicated that this was an
important reason for choosing to take part in this project. Despite
the lack of prior experience, nearly all of the students enjoyed the
course, even if they found it to be quite challenging. Many students
would have liked to have had more time. They would have liked to get a
better understanding of what they were doing and to be able to do more
programming.

Even though the web application and interpreter were available for
exploring Prolog, the students indicated that they were glad that they
had the opportunity to ask questions in class.

The spring 2012 class had more extensive lecture notes to work with, which
also contained several exercises. The students whom we spoke to indicated that
this greatly helped them in understanding the theory. Half of those students
also indicated that taking part in this course made them strongly consider
choosing a Computer Science education. 

\paragraph{Our experience}

Having run the course for two years now, we would like to reflect further on
our experience. First of all, is Prolog a good choice? In the introduction we
motivated our choice for Prolog: it is a simple language with interesting
semantics, that we expect very few students to have studied already. In our
experience, students enjoyed learning Prolog. They get started almost
immediately as the syntax is extremely lightweight. Students can get frustrated
at times---many are not used to the precision necessary to write computer
programs. Missing parentheses are an all too common mistake. Perhaps better IDE
support or a structure editor would help.

Students enjoy learning to program. They enjoy the intellectual challenge and
seeing their code work. Unfortunately, once they have finished their projects,
we think it is unlikely they will continue programming in Prolog. One student
explicitly asked: what language do I learn next? This is a clear drawback of
our choice for Prolog. Teaching a language such as JavaScript, that they have
running in their browser already, would increase the chance that the students
continue programming at home. The downside of using JavaScript, however, is the
higher barrier to entry. We would need to spend more time explaining syntax,
for instance. Perhaps the Racket approach would work well here: starting with a
simple language (fragment) that is incrementally extended to a general purpose
programming language.

One reviewer asked if it was too easy to guess the solution to the Prolog
exercises. This was not our experience at all! In the interactive proof
assistant sessions, where students need to assemble a derivation, this may be
the case. The students can try various rules until they find one that fits.
Even if they do produce derivations by trial and error, we find that seeing the
unification happening before their eyes can still be instructive. The Prolog
predicates we ask them to write are much harder to get right by trial and
error. As soon as any recursion is involved, we believe the chances of guessing
the right solution are very low.

We aimed to teach the students programming and Computer Science. How much do
they really learn in the limited time we have available? After completing the
course, the students have written a small program that computes something
interesting starting from scratch. In a sense, they have learned to program. Of
course, they have plenty to learn before they can write any kind of
application, but we feel that all the students take their first steps in
computer programming, however small these may be. Did the students learn about
Computer Science? Yes and no. As we mentioned above, all the students finish
the course with a much clearer idea of what Computer Science is. They are
exposed to concepts such as algorithms and search trees, but they would need
further practice to understand these abstract concepts completely. For example,
we believe that most students could visit a tree in depth-first order, but very
few could give an accurate pseudo code implementation of depth-first search.

Finally, we learned the importance of setting up a stable teaching environment.
Despite having tested our web application before the project started, students
still managed to break things in unexpected ways. Having SWI Prolog installed
on the machines was a good fall-back option.

\subsection*{Related Work}

There are several similar initiatives aimed at teaching programming to
secondary school students. Bootstrap~\cite{bootstrap} is a curriculum aimed at
middle-school children of ages 11 to 14. It focuses on teaching kids how to
program video games using algebraic and geometric concepts that they know from
their mathematics classes.  As such, the curriculum is designed to work in
conjunction with the regular middle-school program. It consists of nine units
and some supplemental lessons. Students work with the functional programming
languages Racket and Scheme. It also offers a web-based environment in which
the children can write and execute code and see the result appear in the web
browser.

Haskell for Kids~\cite{smith2011} targets a similar audience as Bootstrap,
namely children around the age of 12 or 13. It teaches the basics of Haskell
programming by letting the children draw images and even animations using the
Gloss library~\cite{lippmeier2010}. Similarly to bootstrap, Haskell for Kids
offers a web-based development interface.

``How to Design Programs''~\cite{felleisen2003} is a book with content similar
to our curriculum. It teaches Scheme as a core subject for a liberal arts
education. It differs not only in the target audience, but also in focus: it
teaches the student to write programs, with underlying concepts as an aside,
whereas we focus on teaching the student about the underlying concepts with
learning how to program as an aside.

Khan Academy~\cite{khan} offers a collection of introductory video lectures on
Python. Exercises are not included. The target audience is not specified, but
the level of the videos suggests that it is aimed at teenagers and above.

Alice~\cite{conway1997} takes a more graphical approach to teaching
programming. It focusses on visual learners by enabling them to drag and drop
objects into a program to create an animation or a game. As opposed to the
previous works, it uses a Java-like language, wrapped in a graphical user
interface of a custom IDE. It is targeted at post-secondary education students.

Scratch~\cite{maloney2004} provides a drag and drop interface to programming in
a custom IDE. The language itself is imperative, with an event-driven
concurrency model. The language has a simple, but intuitive type system that
shows types as different shapes. Scratch is aimed at children between 8 and 16
years of age.

Microsoft have developed Small Basic~\cite{smallbasic} to encourage people to
learn how to program. Small Basic is a simplified Basic dialect. The IDE is
simple and intuitive, and focuses on providing information about the Small
Basic libraries API. It has a similar target audience to Khan Academy.

\subsection*{Conclusion}
The module we teach is aimed at secondary school students. Besides teaching
them a bit about computer programming, we aim to teach them about
\emph{Computer Science}. We expose them to concepts such as unification and
substitution; explain search trees and search algorithms; and teach them to
solve problems using recursion. We then proceed to show them how these concepts
are implemented in an interpreter by showing an implementation in Haskell. We
feel that this really distinguishes our course from many of the existing
approaches already available.

\bibliographystyle{eptcs}
\bibliography{bibliography.bib}
\end{document}